\documentclass{osa-article}

\journal{oe}
\articletype{Research Article}
\begin{document}

\title{Tailoring anisotropic absorption in borophene-based structure via critical coupling}

\author{Tingting Liu,\authormark{1} Chaobiao Zhou,\authormark{2} and Shuyuan Xiao\authormark{3,4,*}}

\address{\authormark{1}Laboratory of Millimeter Wave and Terahertz Technology, School of Physics and Electronics Information, Hubei University of Education, Wuhan 430205, China\\
\authormark{2}College of Mechanical and electrical engineering, Guizhou Minzu University, Guiyang 550025, China\\
\authormark{3}Institute for Advanced Study, Nanchang University, Nanchang 330031, China\\
\authormark{4}Jiangxi Key Laboratory for Microscale Interdisciplinary Study, Nanchang University, Nanchang 330031, China\\}

\email{\authormark{*}syxiao@ncu.edu.cn}

%\email{\authormark{*}opex@osa.org} %% email address is required

% \homepage{http:...} %% author's URL, if desired

%%%%%%%%%%%%%%%%%%% abstract %%%%%%%%%%%%%%%%
%% [use \begin{abstract*}...\end{abstract*} if exempt from copyright]

\begin{abstract}
The research of two-dimensional (2D) materials with atomic-scale thicknesses and unique optical properties has become a frontier in photonics and electronics. Borophene, a newly reported 2D material provides a novel building block for nanoscale materials and devices. We present a simple borophene-based absorption structure to boost the light-borophene interaction via critical coupling in the visible wavelengths. The proposed structure consists of borophene monolayer deposited on a photonic crystal slab backed with a metallic mirror. The numerical simulations and theoretical analysis show that the light absorption of the structure can be remarkably enhanced as high as 99.80$\%$ via critical coupling mechanism with guided resonance, and the polarization-dependent absorption behaviors are demonstrated due to the strong anisotropy of borophene. We also examine the tunability of the absorption behaviors by adjusting carrier density and lifetime of borophene, air hole radius in the slab, the incident angle and polarization angle. The proposed absorption structure provides novel access to the flexible and effective manipulation of light-borophene interactions in the visible, and shows a good prospect for the future borophene-based electronic and photonic devices. 
\end{abstract}

%%%%%%%%%%%%%%%%%%%%%%%%%%  body  %%%%%%%%%%%%%%%%%%%%%%%%%%
\section{\label{sec:1}Introduction}
Recent decades have witnessed the rapidly growing interests in two-dimensional (2D) materials, a new family of nano-materials with peculiar band structures and unique optical properties\cite{Xia2014}. The development of 2D materials has enabled effective manipulation of incident light over a wide wavelength scale with the promise to achieve next-generation nanophotonic devices such as optical modulators, field effect transistors, photodetectors, and biosensors\cite{Gan2013,Lopez-Sanchez2013,Roy2014,Menard-Moyon2020,Cheng2020}. To boost the light-matter interaction limited by their monolayer nature, one intuitive solution is to excite the plasmonic response of the 2D materials, for instance, graphene plasmons in the terahertz and infrared and the localized anisotropic surface plasmon resonance of nanostructured black phosphorous in the mid-infrared\cite{Grigorenko2012,Zhu2013,Low2014,Liu2016,Lu2017,Cai2019,Xia2020}. Alternatively, the interactions can also be enhanced by integrating 2D materials with the resonant structures such as plasmonic nanoantennas, Fabry-Perot cavity, hyperbolic metamaterials, et al. \cite{Fang2012,Ferreira2012,Butun2015,Xia2017,Lu2017a,Wang2017,Xiao2019}. Among them, the critical coupling mechanism becomes an excellent candidate for absorption enhancement of 2D materials due to the simple design, the remarkable field confinement, and the flexible tunability merits. This method has been theoretically and experimentally employed in the whole 2D material family, providing a good way to improve the light-matter interaction of 2D materials and holding great potential for the development of novel functional devices with superior performance\cite{Piper2014,Liu2014,Huang2016,Guo2016,Jiang2017,Fan2017,Li2017,Li2018,Qing2018,Wang2019,Liu2019,Xiao2020,Wang2020a,Wang2020,Liu2021}.

Very recently, borophene, the two-dimensional boron polymorphs, has become a new member of the wonderful 2D material family and provides a novel building block for nanoscale materials and devices. In contrast with the semiconducting nature of the bulk boron, borophene sheet is predicted to be metallic with high electron density, which has been demonstrated by the atomic-scale characterization of the synthesized sheets on silver surfaces\cite{Yang2008,Wu2012,Mannix2015,Feng2016,Feng2017}. Such 2D metal characteristic is complementary to those of the existing 2D materials only accessible to 2D semi-metal as graphene or 2D semiconductor as MoS$_2$. Moreover, similar with black phosphorus, borophene shows highly anisotropic electronic properties and leads to interesting physical phenomena\cite{Huang2017,Liu2019a}. The high electron density and strong anisotropy of borophene have motivated the further investigation of manipulating light-matter interaction. Inspired by the plasmon modes in other existing 2D materials, the patterned borophene is demonstrated to exhibit anisotropic plasmonic behavior in visible wavelengths\cite{Dereshgi2020}. However, the nanostructured borophene in the isolated fashion is usually involved with complicated fabrication techniques. For practical applications, it is highly desirable to exploit the borophene monolayer form for the interaction enhancement. 

Following this idea, we propose a borophene-based absorption structure via critical coupling in the visible range, where borophene monolayer is covered on top of a photonic crystal slab backed by a metallic mirror. We show that by critical coupling with guided resonance, the light-borophene interaction is highly enhanced with the total absorption up to 99.80$\%$. The absorption behaviors can be tailored by tuning the carrier density and carrier lifetime of borophene, the air hole radius of the slab, and the incident light angles. Specially, owing to the strong anisotropy of borophene, the structure exhibits distinct absorption spectra under TM and TE polarizations. The proposed structure presents a promising way to enhance the interaction of light in borophene, which may promote the development of the novel borophene-based nanodevices.

\section{\label{sec:2}Structure design and numerical model}
On the periodic table, boron is the neighbor of carbon and they have the same short covalent radius and the flexibility to adopt $sp^{2}$ hybridization, which is favorable to form the low-dimensional borophene. As boron has one less electron than carbon, the monolayer structure needs to be stabilized by balancing out the two-center bonding in the hexagonal regions and three-center bonding in the triangular\cite{Tang2007,Tang2010}. Depending on the connectivity of the boron, various polymorphs of borophene have been theoretically predicted and confirmed by experimental synthesis, such as the lowest-energy monolayer structure by Mannix et al., the $\beta_{12}$ and $\chi_{3}$ polymorphs by Feng et al. on Ag(111) surface using molecular beam epitaxy\cite{Mannix2015,Feng2016,Feng2017}. In particular, instead of the substrate-supported ultrahigh-vacuum growth techniques, the synthesis of freestanding atomic sheets of borophene has been demonstrated through the combination of the modified Hummer’s technique (chemical exfoliation) and sonochemical exfoliation by Ranjan et al.\cite{Ranjan2019,Ranjan2020}, which largely expands the applications in nanophotonics. Since our main aim is to explore the total absorption of borophene via critical coupling as other 2D materials, the borophene monolayer with high density of electrons and strong anisotropy is adopted as the building block in this work, where its schematic could be found in Fig. 1(a) and (b). In the visible wavelengths, the surface conductivity of the borophene monolayer can be modeled using a simple semiclassical Drude model as\cite{Dereshgi2020}
\begin{equation}
	\label{eq:1}
	\sigma_{jj}=\frac{iD_j}{\pi(\omega+\frac{i}{\tau})}, D_j=\frac{\pi e^2 n_s}{m_j},
\end{equation}
where $j$ is the direction concerned that is taken to be $x$ or $y$ in this study, $\omega$ is the incident light frequency, $\tau$ is the mean free time of electron ranging from 10 fs to 65 fs, $e$ and $n_s$ represent electron charge and free carrier density, respectively, $D_j$ is the Drude weight, $m_j$ is the effective electron mass along different directions, and $m_x=1.4m_0$, $m_y=5.2m_0$ with $m_0$ is the rest mass of electron. The effective permittivity of the borophene monolayer can be derived from the surface conductivity along each direction as 
\begin{equation}
	\label{eq:2}
	\varepsilon_{jj}=\varepsilon_{r}+\frac{i\sigma_{jj}}{\varepsilon_{0}\omega d_B},
\end{equation}
where $\varepsilon_{r}=11$ is the relative permittivity of boron, $\varepsilon_{0}$ is the permittivity of free space, and $d_B$ is the thickness of borophene. Here the assumption of the thickness $d_B=0.3$ nm is adopted to simulate borophene monolayer and it can resemble the real results because it is small enough compared with the wavelength of interest\cite{Mannix2015,Dereshgi2020}. With the thickness $d_B=0.3$ nm, the polarization-dependent permittivity of borophene is displayed in Fig. 1(c) and (d), exhibiting the strong anisotropic optical behaviors. 

\begin{figure}[htbp]
	\centering
	\includegraphics% Here is how to import EPS art
	[scale=0.50]{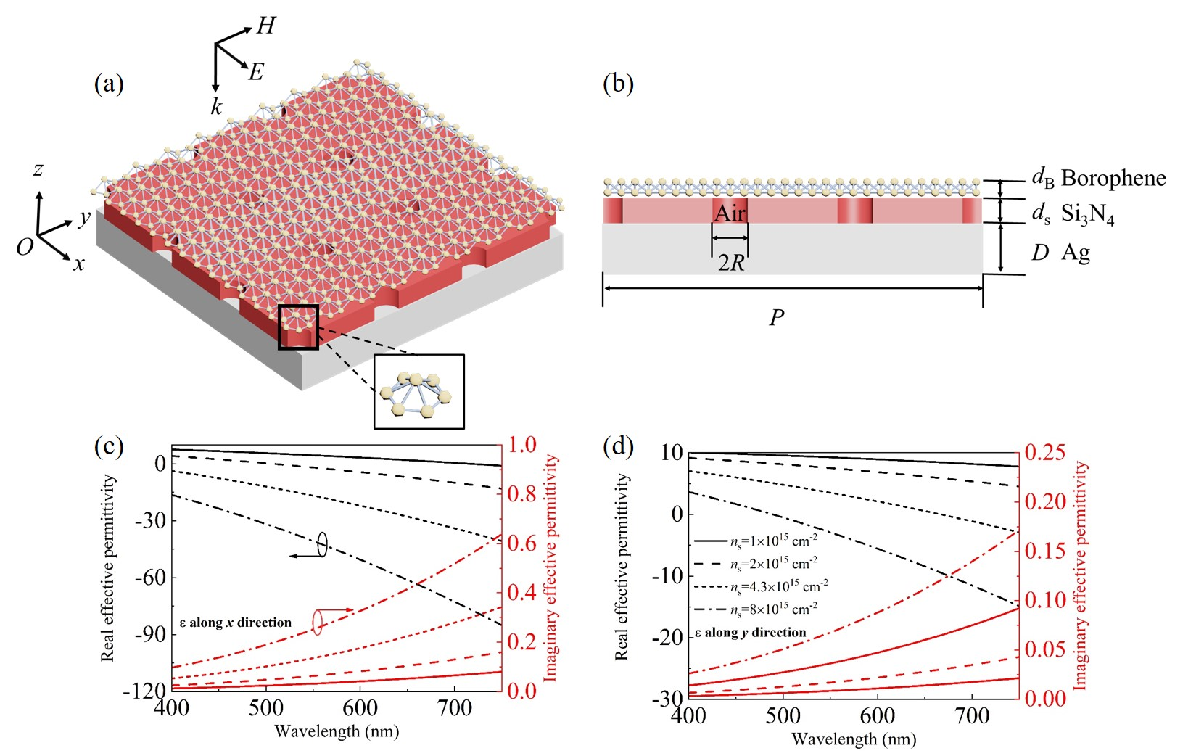}
	\caption{\label{fig:1} (a) Schematic illustration of the proposed borophene-based absorption structure. The enlarged illustration represents the atomic structure of the borophene monolayer. (b) Cross-sectional view of the proposed structure. The lattice period is $P$, and the air hole radius is $R$. The thicknesses of borophene monolayer, Si$_{3}$N$_{4}$ photonic crystal slab and silver layer are denoted by $d_B$, $d_s$, and $D$, respectively. The polarization-dependent effective permittivity of borophene monolayer along (c) $x$ direction and (d) $y$ direction. The black and red lines represent the real and imaginary parts, respectively. }
\end{figure}

In the proposed one-port coupling configuration, borophene monolayer is deposited on top of a photonic crystal slab backed with a metallic mirror, as illustrated in Fig. 1 (a) and (b). In order to achieve critical coupling absorption in the visible, the lossless photonic crystal slab made from Si$_{3}$N$_{4}$ with periodic air holes is adopted as the resonator, where the correct choice of the in-plane periodicity enables the phase-matched coupling between guided mode and the external radiation, giving rise to the excitation of the guided resonance and strongly confined electromagnetic field. The placement of the borophene monolayer on the top introduces the lossy thin film, while it shows little impact on the field distribution of resonance inside the slab. The metallic mirror is employed to reflect light and suppress the transmission. We perform numerical simulations using the finite-difference time-domain (FDTD) method within the wavelengths between 530-610 nm, while the refractive index of $n=2$ is used for Si$_{3}$N$_{4}$ slab and the permittivity of the silver mirror within these wavelengths is given by Drude model with the plasmon frequency $\omega_p=1.38\times10^{16}$ rad/s and the collision frequency $\gamma_p=2.73\times10^{13}$ rad/s \cite{Ordal1985}. In the simulations, the periodic boundary conditions are utilized in both $x$ and $y$ directions and the perfectly matched layer is applied along the $z$ direction since the plane waves are normally incident from -$z$ direction. The mesh inside the borophene monolayer is 0.15 nm that is fine enough to resemble the real results. Then the absorption of the proposed structure is simplified as $A=1-R$ because of the blocked transmission from the metallic mirror.

\section{\label{sec:3}Results and discussions}
In the initial simulations, the borophene monolayer is set with the free carrier density $n_s=4.3$$\times$10$^{15}$ cm$^{-2}$ and the carrier lifetime $\tau$=60 fs, while the structural parameters are optimized to achieve critical coupling condition with the values of the lattice period $P= 500$ nm, the Si$_{3}$N$_{4}$ thickness $d_s=100$ nm, the Ag thickness $D=200$ nm, the air hole radius $R=70$ nm, respectively. As shown in Fig. 2(a) and (b), the simulated absorption spectra under normally incidence are depicted. Due to the strong anisotropic behavior of borophene, the absorption spectra exhibit direction dependence for TM and TE polarizations. In Fig. 2(a) for TM polarization, the incident light is completely absorbed with maximum absorption amplitude up to $99.80\%$ at the resonance wavelength of 565.09 nm, while the spectrum in Fig. 2(b) for TE polarization shows a peak absorption of $92.13\%$ at 571.56 nm. It is also observed that the absorption of TM polarization is only $0.87\%$ at the resonance wavelength of TE polarization. 

\begin{figure}[htbp]
	\centering
	\includegraphics% Here is how to import EPS art
	[scale=0.50]{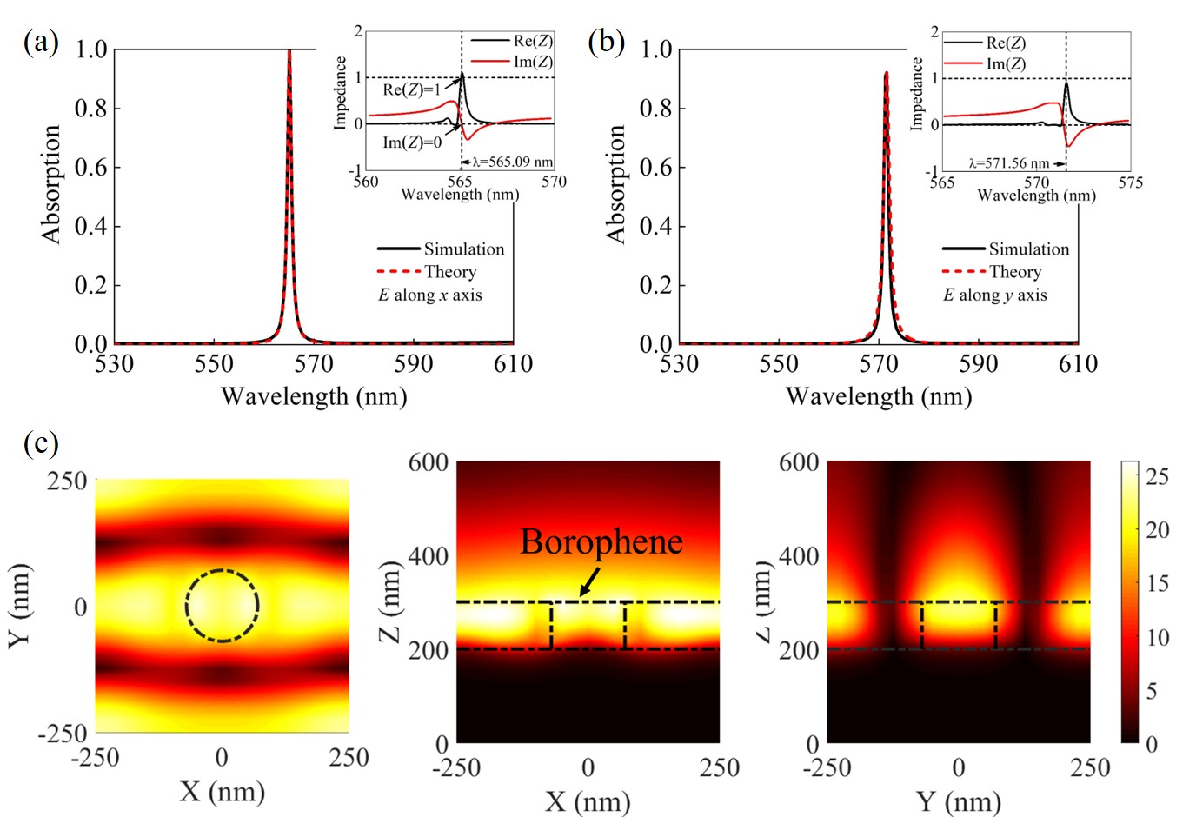}
	\caption{\label{fig:2} Simulated and theoretical absorption spectra of the proposed structure  for (a) TM polarization and (b) TE polarization. The effective impedances of the corresponding absorption spectra in the vicinity of the resonance are shown in the insets. (c) Electric field distributions |E| for TM polarization at the resonance wavelength. }
\end{figure}

To obtain a comprehensive understanding of the absorption characteristics, the coupled mode theory (CMT) is utilized to describe the input-output properties of the resonant system. Here only the lowest-order guided resonance among the wavelengths of interest is considered. When the incident light is coupled to the guided mode, the electromagnetic field near the borophene monolayer is significantly enhanced in the vicinity of the wavelength of the guided resonance, which remarkably boosts the light-matter interaction, and the light absorption of the structure could be greatly improved. In presence of the external leakage rate $\gamma$ and the intrinsic loss rate $\delta$ in the system, the reflection coefficient can be described as\cite{Piper2014}
\begin{equation}
	\label{eq:3}
	\Gamma=\frac{y}{u}=\frac{i(\omega-\omega_{0})+\delta-\gamma}{i(\omega-\omega_{0})+\delta+\gamma},
\end{equation}
where $y$ and $u$ represent output and input wave amplitudes, $\omega_{0}$ is the resonance frequency. The absorption is calculated as 
\begin{equation}
	\label{eq:4}
	A=1-|\Gamma|^{2}=\frac{4\delta\gamma}{(\omega-\omega_{0})^{2}+(\delta+\gamma)^{2}}.
\end{equation}
From the equations, the critical coupling condition can be achieved when the external leakage and intrinsic loss rates are the same as $\delta=\gamma$ at the resonant frequency of $\omega_{0}$. That is to say, the reflection of the system will vanish and all the incident light will be totally absorbed under this condition. At the same time, the effective impedance of the whole structure is supposed to equal with that of the free space, i.e. $Z=Z_0=1$. The effective impedance of the one-port configuration is given by\cite{Smith2005,Szabo2010}
\begin{equation}
	\label{eq:5}
	Z=\frac{(T_{22}-T_{11})\pm\sqrt{(T_{22}-T_{11})^{2}+4T_{12}T_{21}}}{2T_{21}},
\end{equation}
where $T_{11}$, $T_{12}$, $T_{21}$, and $T_{22}$ are the elements of the transfer matrix of the structure calculated from the scattering matrix elements as following, and the two roots of $Z$ correspond to the two paths of light propagation with plus sign denoting the positive direction.
\begin{equation}
\label{eq:6}
T_{11}=\frac{(1+S_{11})(1-S_{22})+S_{21}S_{12}}{2S_{21}},
\end{equation}
\begin{equation}
\label{eq:7}
T_{12}=\frac{(1+S_{11})(1+S_{22})-S_{21}S_{12}}{2S_{21}},
\end{equation}
\begin{equation}
\label{eq:8}
T_{21}=\frac{(1-S_{11})(1-S_{22})-S_{21}S_{12}}{2S_{21}},
\end{equation}
\begin{equation}
\label{eq:9}
T_{22}=\frac{(1-S_{11})(1+S_{22})+S_{21}S_{12}}{2S_{21}}.
\end{equation}

Based on CMT, the theoretical absorption spectra are provided and depicted as the dashed lines in Fig. 2(a) and (b). By fitting the theoretical and simulated curves, the intrinsic loss and external leakage rates can be obtained as $\delta=\gamma=1.45$ THz for TM polarization. Then the theoretical quality factor $Q_{CMT}$ of the resonant system is calculated as 1147.27 by a simplified form $Q_{CMT}=\omega_{0}/4\delta$. In comparison, the quality factor $Q_0$ of the simulated spectrum is obtained as 1146.24 by the definition of $Q_0=\lambda_0/ \Delta\lambda_0$ where $\Delta\lambda_0=0.49$ nm is the full width at half maximum (FWHM) of the spectrum. The theoretical and simulated quality factor show nearly identical values, suggesting that the total absorption of the system can be attributed to critical coupling. In Fig. 2(c), the electric field distribution $|E|$ at the resonance wavelength of TM polarization also demonstrates that the strong field confinement of the guided resonance near the borophene monolayer results in the remarkable absorption enhancement of the structure. In contrast, the two rates for TE polarization are fitted as $\delta=0.82$ THz and $\gamma=1.45$ THz. As the polarization direction changes, the intrinsic loss decreases due to the imaginary part of the effective permittivity of borophene shows reduced values along $y$ direction relative to the values along $x$ direction. During this process, the leakage rate of the system keeps unchanged. Hence, the system is in the state of over coupling with $\delta<\gamma$ for TE polarization. From the point of impedance matching, the total absorption is achieved when the effective impedance of the structure matches with that of free space, which is the case for TM polarization with $Z=1.09-i0.04$. The mismatch between the impedance $Z=1.36-i0.22$ of TE polarization and that of free space accounts for the nonperfect absorption.  

\begin{figure}[htbp]
	\centering
	\includegraphics% Here is how to import EPS art
	[scale=0.50]{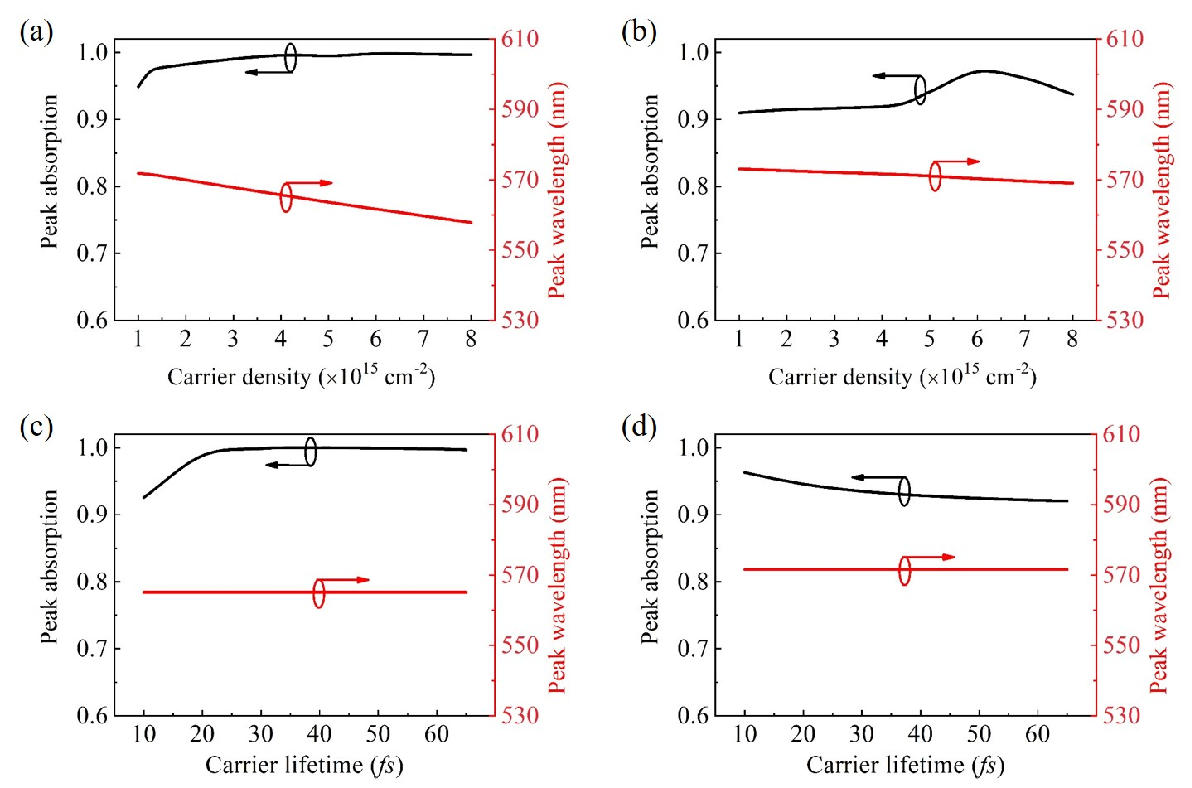}
	\caption{\label{fig:3} Variations of the absorption peaks and resonance wavelengths at different borophene properties: carrier density for (a) TM polarization and (b) TE polarization, and carrier lifetime for (c) TM polarization and (d) TE polarization. }
\end{figure}

Given the fact that the perfect absorption is achieved by manipulating the critical coupling condition, the absorption behaviors can be actively tailored by adjusting the intrinsic loss and external leakage rates of the proposed structure. In the wavelength of interest, the borophene monolayer plays the essential role of lossy thin film and contributes to the intrinsic loss in the coupled system. This motivates us to investigate the influence of the borophene properties on the absorption behaviors under different polarizations. In Fig. 3, the dependences of the absorption peak and the resonance wavelength of spectra on the free carrier density $n_s$ and the carrier lifetime $\tau$ in borophene are depicted. For TM polarization in Fig. 3(a), the absorption peak shows an increases from $94.88\%$ to $99.80\%$ as the carrier density $n_s$ reaches $4.3\times$10$^{15}$ cm$^{-2}$, and then the system keeps in critical coupling state with maximum absorption despite of the further increase of $n_s$. In Fig. 3(b), the absorption peak for TE polarization is unchanged firstly and then shows a maximum of $97.11\%$ at $n_s= 6\times$10$^{15}$ cm$^{-2}$. The resonance wavelengths of both polarizations show linear blue shift owing to the smaller real part of the effective permittivity of borophene as $n_s$ increases, implying the feasibility within a broad wavelength range. In Fig. 3(c) and (d), as the carrier lifetime $\tau$ varies within the empirical range from 10 fs to 65 fs, the absorption peak increases at first and keeps at the maximum of $99.80\%$ for TM polarizaiton, while it decreases slightly from $96.30\%$ to $92.05\%$ for TE polarization. It is also observed that the variations of $\tau$ show no impact on the resonance wavelength, which can be explained by its little influence on the real part of the borophene permittivity. Moreover, all the absorption peaks in Fig. 3 show large amplitudes with above $90\%$, revealing the robustness of the proposed absorption structure.

In addition to the modulation of the instinct loss resulting from borophene properties, the absorption behaviors can also be controlled by the variations of the external leakage rate. In the guided resonance system supported by the photonic crystal slab, the external leakage rate $\gamma$ is mainly controlled by the ratio between the air hole radius and lattice period, i.e. $R/P$ \cite{Piper2014}. Here we investigate the variations of absorption peak and resonance wavelength with respect to the air hole radius while other parameters are fixed, as shown in Fig. 4(a) and (b). Obviously, the absorption behaviors show more sensitive dependence on this geometrical parameter than that on borophene properties. In detail, for TM polarization, the absorption peak exhibits significant increase from $21.83\%$ to $99.80\%$ as radius varies from 50 nm to 70 nm and then shows decline tendency. As radius increases, the external leakage rate also increases from 0.09 THz to 5.5 THz while the instinct loss rate keeps almost unchanged as 1.45 THz for TM polarization, and this rule applies to the rates for TE polarization. As a result, the coupled system evolves from the state of under coupling, through critical coupling and to over coupling. Due to the reduced effective refractive index of the guided resonance with the larger radius, the resonance wavelength of the absorption spectra displays an obvious blue shift from 572.83 nm to 556.61 nm. The influence of air hole radius on the absorption behaviors for TM polarization also applies to that of TE polarizations. Thus, the leakage rate could be engineered by adjusting geometrical parameters, leading to the tunable absorption in borophene-based structure. 

\begin{figure}[htbp]
	\centering
	\includegraphics% Here is how to import EPS art
	[scale=0.50]{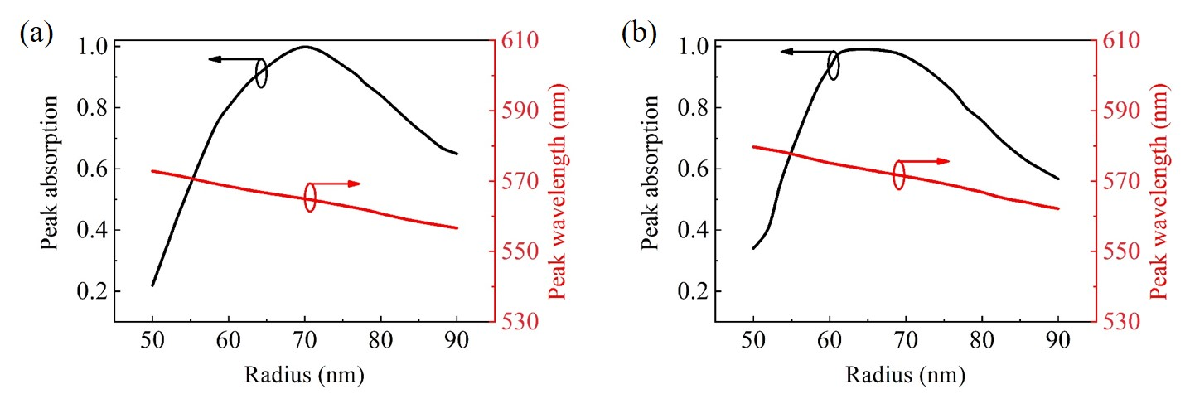}
	\caption{\label{fig:4} Variations of the absorption peaks and resonance wavelengths at different air hole radius in photonic crystal slab for (a) TM polarization and (b) TE polarization. }
\end{figure}

Next, the dependence of the absorption behaviors on the angular dispersion is investigated. Different absorption spectra of TM and TE polarizations are exhibited under oblique incidence, as illustrated in Fig. 5(a) and (b), respectively. When the incident angles varies from 0$^{\circ}$ to 6$^{\circ}$, the absorption peak for TM polarization remains almost unchanged with the amplitudes above 95$\%$, accompanied by a quite slight blue shift of resonance wavelength from 565.09 nm to 564.36 nm. In contrast with the tolerant behavior of TM polarization, the absorption behaviors for TE polarization is more sensitive to the incident angles. It can be seen that the two distinct absorption peaks appear under oblique incidence because of the excitation of another resonant mode. In Fig. 5(c), the absorption behaviors as function of the polarization angles is provided. As the polarization angle is 0$^{\circ}$, the total absorption for TM polarization is observed with amplitude of $99.80\%$ at resonance wavelength of 565.09 nm. When the polarization angle varies from 0$^{\circ}$ to 90$^{\circ}$, there is a gradual decrease in the absorption peaks for TM polarization at 565.09 nm and a continuous increase in the absorption peak for TE polarization at 571.57 nm. Finally, the peak for TM polarization disappears and only the one for TE polarization exists with its maximum. The resonance wavelength of the peaks remains intact during the variations, suggesting no coupling effect between the guided modes. 

\begin{figure}[htbp]
	\centering
	\includegraphics% Here is how to import EPS art
	[scale=0.60]{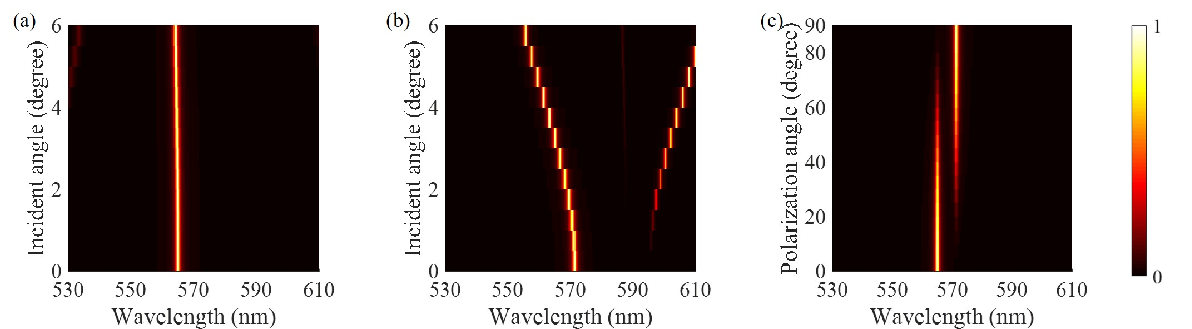}
	\caption{\label{fig:5}  Variations of the absorption behaviors at different incident angles for (a) TM polarization and (b) TE polarization, and (c) at different polarization angles under normal incidence. }
\end{figure}

\section{\label{sec:4}Conclusions}
In conclusion, we propose and theoretically demonstrate the enhanced light-matter interaction in the borophene monolayer via critical coupling in the visible wavelengths. In the proposed simple compact absorption structure consisting of borophene covered on a photonic crystal slab backed with metallic mirror, the numerical results show that the absorption can be remarkably enhanced up to $99.80\%$ at normal incidence. The physical regime of the total absorption is the field confinement through the principle of critical coupling with guided resonance. The proposed structure exhibits polarization-dependent absorption spectra along the $x$ and $y$ directions owing to the anisotropic nature of borophene. We also examine various parameters such as borophene properties including carrier density and lifetime, geometrical parameters including air hole radius in the slab, the angular dispersion including the incident angles and polarization angles to analyze the tunablity of the absorption behaviors. The proposed structure opens a new path to improve the light-borophene interaction in the visible for future borophene-based electronic and photonic devices. 

\begin{backmatter}
\bmsection{Funding}
National Natural Science Foundation of China (11847132, 11947065, 61901164, 12004084); Natural Science Foundation of Jiangxi Province (20202BAB211007); Interdisciplinary Innovation Fund of Nanchang University (2019-9166-27060003); Natural Science Research Project of Guizhou Minzu University (Grant No. GZMU[2019]YB22); China Scholarship Council (202008420045).

\bmsection{Acknowledgments}
The authors would also like to thank Dr. X. Jiang for beneficial discussions on the critical coupling mechanism.

\bmsection{Disclosures}
The authors declare no conflicts of interest.

\bmsection{Data availability} Data underlying the results presented in this paper are not publicly available at this time but may be obtained from the authors upon reasonable request.

\end{backmatter}

%%%%%%%%%%%%%%%%%%%%%%% References %%%%%%%%%%%%%%%%%%%%%%%%%
\bibliography{Ref}

%%%%%%%%%% If preparing manually:
% \begin{thebibliography}{1}
% \newcommand{\enquote}[1]{``#1''}

% \bibitem{Zhang:14}
% Y.~Zhang, S.~Qiao, L.~Sun, Q.~W. Shi, W.~Huang, L.~Li, and Z.~Yang,
%   \enquote{Photoinduced active terahertz metamaterials with nanostructured
%   vanadium dioxide film deposited by sol-gel method,}
%   {\protect\JournalTitle{Optics Express}} \textbf{22}, 11070--11078 (2014).

% \bibitem{OSA}
% {Optical Society}, \enquote{{OSA Publishing},}
%   \url{http://www.osapublishing.org}.

% \bibitem{FORSTER2007}
% P.~Forster, V.~Ramaswamy, P.~Artaxo, T.~Bernsten, R.~Betts, D.~Fahey,
%   J.~Haywood, J.~Lean, D.~Lowe, G.~Myhre, J.~Nganga, R.~Prinn, G.~Raga,
%   M.~Schulz, and R.~V. Dorland, \enquote{Changes in atmospheric consituents and
%   in radiative forcing,} in \enquote{Climate Change 2007: The Physical Science
%   Basis. Contribution of Working Group 1 to the Fourth assesment report of
%   Intergovernmental Panel on Climate Change,}  S.~Solomon, D.~Qin, M.~Manning,
%   Z.~Chen, M.~Marquis, K.~B. Averyt, M.~Tignor, and H.~L. Miler, eds.
%   (Cambridge University Press, 2007).

% \end{thebibliography}

\end{document}